# Comparison Based Analysis of Different Cryptographic and Encryption Techniques Using Message Authentication Code (MAC) in Wireless Sensor Networks (WSN)

Sadaqat Ur Rehman*, Muhammad Bilal**, Basharat Ahmad**, Khawaja Muhammad Yahya**, Anees Ullah**, Obaid Ur Rehman*

*Department of Electrical Engineering, Sarhad University of Science and Technology, Peshawar, 25000, Pakistan.
**Department of Computer Systems Engineering, N-W.F.P. University of Engineering & Technology Peshawar, 25000, Pakistan.

## Abstract

Wireless Sensor Networks (WSN) are becoming popular day by day, however one of the main issue in WSN is its limited resources. We have to look to the resources to create Message Authentication Code (MAC) keeping in mind the feasibility of technique used for the sensor network at hand. This research work investigates different cryptographic techniques such as symmetric key cryptography and asymmetric key cryptography. Furthermore, it compares different encryption techniques such as stream cipher (RC4), block cipher (RC2, RC5, RC6 etc) and hashing techniques (MD2, MD4, MD5, SHA, SHA1 etc). The result of our work provides efficient techniques for communicating device, by selecting different comparison matrices i.e. energy consumption, processing time, memory and expenses that satisfies both the security and restricted resources in WSN environment to create MAC.

## 1. Introduction

Wireless sensor networks (WSN) have the advantage over traditional networks in many ways such as large scale, autonomous nature and dense deployment [1]. Moreover, it has increased fault tolerance because if a sensor node fails others can collect/process data. Because of its ad-hoc nature it becomes more attractive in certain applications such as military, environmental observation, syndrome surveillance, supply chain management, fire detection, vision enabling, energy automation, building administration, gaming, health and other commercial and home applications [2], [3], [4], [5], [6], [7], [8], [9], [10], [11], [12], [13], [14], [15].

With the wide deployment of WSN for multi-faceted applications security is becoming a growing concern. For example, in a battlefield, a military communication network used for sensitive information interchange can be hacked by its adversaries if the WSN has security holes causing severe loss of life and machinery. Similarly, many social problems can be created if personal information flowing on health care systems is intercepted [16]. Security of WSN is a big challenge due to its limited resources such as energy, power supplies, small memory, computation and communication capabilities [17], [18], [19], [20], [1]. This is the reason that traditional security techniques cannot be applied on sensor networks, indirectly rising the need to make sensor network economically feasible [21], [22].

Cryptographic algorithm plays an important role in the security and resource conservation of wireless sensor networks (WSN) [23], [24]. This work spotlights different cryptographic techniques and compares different encryption techniques such as stream cipher (RC4), block cipher (RC2, RC5, RC6 etc) and hashing techniques (MD2, MD4, MD5, SHA, SHA1 etc). Our main aim of working in this survey paper is to put forward a cryptographic and encryption technique that creates Message Authentication Code (MAC) in wireless sensor networks (WSN), which is more feasible in the restricted resources of wireless sensor networks (WSN) and also provide good security in communication as well.

The rest of the paper is outlined as follows. We present our critical review questions in section II; then we analyze these review questions in section III and finally we deduct our conclusion in section IV.

## 2. Research Questions

The critical review questions we are in quest of to answer are:

- Why we prefer symmetric keys over public keys in WSN?
- Which method and algorithm is best to create MAC in Wireless Sensor Networks?

- Can we apply all given methods to create MAC like block cipher, stream cipher, hash function and unconditional secure?

## 3. Analysis and Discussion

In this section we analyze the result of our research questions. Our questions are:

### 3.1 Why we prefer symmetric keys over public keys in WSN?

Symmetric key technique uses a single key called secret key which uses less mathematics, results in less computation, on the other hand asymmetric key technique uses both public and private keys, results in more processing and consumes more energy. Symmetric key techniques offer better energy efficiency as compared to public key that is why most researches use it for creating MAC in WSN.

According to [25] public key is used in some applications for secure communications e.g. SSL (Secure Socket Layer) and IPsec standards both use it for their key agreement protocols. But it consumes more energy and also it is more expensive as compared to symmetric key.

[26] has given a reason that public key consumes more energy due to great deal of computation and processing involved, which makes it more energy consumptive as compared to symmetric key technique e.g. a single public key operation can consume same amount of time and energy as encrypting tens of megabits using a secret key cipher.

According to [27], the more consumption of computational resources of public key techniques is due to the fact that it uses two keys. One of which is public and is used for encryption, and every one can encrypt a message with it and other is private on which only decryption takes place and both keys has a mathematical link, the private key can be derived from a public key. In order to protect it from attacker the derivation of private key from the public is made difficult as possible like taking factor of a large number which makes it impossible computationally. Hence, it shows that more computation is involved in asymmetric key technique thus we can say that symmetric key is better to choose for WSN.

According to [28] the cost of public key is much more expensive as compared to symmetric key for instance, *a 64 bit RC5 encryption on ATmega128 8MHz takes 5.6 milliseconds, and a 160 bit SHA1 has function evaluation takes only 7.2 millisecond's*. These symmetric key algorithms are more than 200 times faster than Public key algorithms.

Public key cryptography is not only expensive in computation but also it is more expensive in communication as compared to symmetric key cryptography. According to [4] to send a public key from one node to another, at least 1024 bits required to be sent if the private key is 1024 bits long.

[25], [26], [27] and [28] suggest that symmetric key cryptography is better than asymmetric key cryptography in both cost and computation.

### 3.2 Which method and algorithm is best to create MAC in Wireless Sensor Networks?

According to [29] Block Cipher is more secure as compared to Stream Cipher this is because of the facts:

- Attacks such as differential attacks on block cipher are also applied to stream cipher.
- Attacks such as correlation attacks on stream cipher are not valid on block cipher.
- Algebraic attacks on stream cipher are more effective.
- Guess and set attacks against stream ciphers recover the key or any plaintext.
- Generic time/memory attacks are stronger against stream cipher than block cipher.

Because of these facts it shows that block cipher is more secure as compared to stream cipher and thus the stream cipher will be replaced with block cipher except few applications.

[30] Compared different attacks on Hash function like birthday attack. He uses MD5 algorithm for this attack and finds out that such an attack needs $2^{64}$ blocks (or $2^{73}$ bits) of data for authentication using the same key. If the communication link has the ability to process 1 Gbit/sec it means one need 250,000 years to process the data needed by such an attack. Even according to [30] on software implementation the popular hash function is faster than the block cipher.

As stream cipher uses a key "K" and initialization vector (IV) for encryption making it more vulnerable to retrieve the plaintext in case different packets use the same IV. If initialization vector is long then it will require additional bytes but our aim is to reduce the packet overhead. Thus we follow the principle *"use an encryption scheme that is as robust as possible in the presence of repeated IVs"*. As stream cipher does not follow this principle so the only way is to use block cipher.

Block cipher has different algorithms such as DES, AES, RC5 and Skipjack. The block cipher used for encryption has an extra advantage i.e. *the most efficient MAC algorithm use a block cipher* [31].

After choosing block cipher for creating MAC, we need to choose algorithm in block cipher. DES is very slow when it is implemented in software. Similarly, experimentations show that AES is quite slow. We find that RC5 and Skipjack are more suitable for sensor networks. One can outperform the other on specific hardware platform. For example, on TinySec platform, although RC5 is slightly faster than Skipjack but it uses 104 extra bytes of RAM per key for good performance. Therefore the default block cipher in TinySec is Skipjack [31]. [31] tested the performance of RC5 and Skipjack on Mica2 sensor node to determine the speed of these two 64 bit block cipher. The time to execute cipher operation on the Mica 2 sensor node is shown as in the Table 1.

Table 1: Time to execute cipher operation on the mica2 sensor nodes [31]

| Block Cipher | Time (ms) | Time (byte times) |
|---|---|---|
| RC5 ( C ) | 0.90 | 2.2 |
| Skipjack ( C ) | 0.38 | 0.9 |
| RC5 ( C, assembly) | 0.26 | 0.6 |

[28] has chosen five popular encryption schemes for study which ranges from stream cipher (RC4) and block ciphers (RC5, IDEA) to hashing techniques ( SHA-1, MD5). RC5 was also chosen for Atmega in the Berkeley Motes SPINS Project. RC5 was chosen on this platform because it uses less memory. [28] Also found that hashing techniques requires an order of magnitude higher overhead.

The parameters used in our paper are shown as in Table 2.

Table 2: Encryption schemes and parameters [28]

| Algorithm | Type | Key/Hash | Block |
|---|---|---|---|
| RC4 [2] | Stream | 128 bits | 8 bits |
| IDEA [2] | Block | 128 bits | 64 bits |
| RC5 [1] | Block | 64 bits | 64 bits |
| MD5 [2][3] | 1-way hash | 128 bits | 512 bits |
| SHA1 [4] | 1-way hash | 128 bits | 512 bits |

On hardware platforms [28] evaluates the performance of these different cryptographic algorithms on different processors that ranges from low end i.e. (4 MHz 8 bit Atmel AVR Atmega 103) to high end (400 MHz 32 bit Intel XScale). Which are shown as in Table 3.

Table 3: Hardware platforms [28]

| Platform | Word Size | Clock Frequency | I/D-S |
|---|---|---|---|
| Atmega 103 | 8 bits | 4 MHz | None |
| Atmega 128 | 8 bits | 16 MHz | None |
| M16C/10 | 16 bits | 16 MHz | None |
| SA-1110 | 32 bits | 206 MHz | 16/8 KB |
| PXA250 | 32 bits | 400 MHz | 32/32 KB |
| UltraSparc2 | 64/32 bits | 440 MHz | 16/16 KB |

Experiments have been performed for different values of selected parameters on all these algorithms, architecture and considered platforms. The functional block of all these algorithms i.e. initialization, encryption and decryption was executed 1000 times using the same input and the result was averaged for these execution.

The execution time overhead for each algorithm and for considered platforms on a log scale is shown as in the Figure1. These are also shown in Table 4.

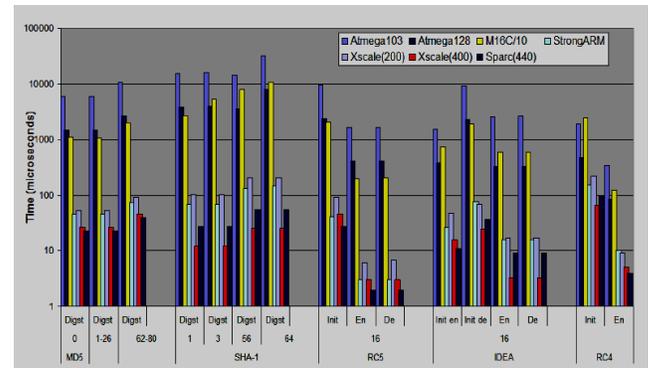

Fig. 1 Execution times [µs] for algorithms, platforms and plaintext sizes [bytes] [28]

Table 4: Execution times [µS] for algorithms, platforms and plaintext sizes [bytes] [28].

| Algorithm | Size | Action | Atmega103 | Atmega128 | M16C/10 | StrongARM | Xscale(400) | Xscale(200) | Sparc(440) |
|---|---|---|---|---|---|---|---|---|---|
| MD5 | 0 | Digest | 5863 | 1466 | 1083 | 46 | 26 | 53 | 23 |
| | 1-26 | Digest | 5890 | 1473 | 1075 | 46 | 26 | 53 | 23 |
| | 62-80 | Digest | 10888 | 2722 | 2011 | 74 | 45 | 90 | 39 |
| SHA-1 | 1 | Digest | 15249 | 3812 | 2651 | 69 | 12 | 102 | 27 |
| | 3 | Digest | 15781 | 3945 | 5303 | 69 | 12.3 | 103 | 27 |
| | 56 | Digest | 14543 | 3636 | 7955 | 133 | 25.8 | 205 | 55 |
| | 64 | Digest | 31107 | 7777 | 10907 | 145 | 25.7 | 207 | 56 |
| RC5 | 16 | Init | 9641 | 2410 | 2074 | 41 | 45 | 91 | 28 |
| | | Enc | 1651 | 413 | 197 | 3 | 3 | 6 | 2 |
| | | Dec | 1636 | 409 | 202 | 3 | 3 | 7 | 2 |
| IDEA | 16 | Init enc | 1523 | 381 | 727 | 26 | 15.54 | 47 | 11 |
| | | Init dec | 9417 | 2354 | 1927 | 76 | 25.16 | 69 | 36 |
| | | Enc | 2555 | 325 | 596 | 16 | 3.24 | 17 | 9 |
| | | Dec | 2614 | 325 | 597 | 16 | 3.27 | 17 | 9 |
| RC4 | 16 | Init | 1886 | 472 | 2455 | 155 | 66.8 | 216 | 96 |
| | | Enc | 344 | 86 | 123 | 10 | 5 | 9 | 4 |

After performing simulation of these algorithms [28] summarizes the result in table V. Comparing the RC4 and RC5 on Atmega 103 shows that the encryption time for both algorithms are close to each other, in fact, RC4 is slightly faster. But, however, by comparing them on Strong ARM, it shows that RC5 is 3 times faster than RC4 algorithm although RC4 operates on 8 bits while RC5 operates on 32 bits.

Comparing RC5 with IDEA on the Atmega 103 showed that RC5 is 1.5 times faster than IDEA. However, both of these algorithms use 64 bit blocks. Hashing techniques needs almost an order of a magnitude higher overhead. Thus RC5 is faster compared to other algorithms like RC4 and IDEA so it requires less processing time and thus less energy consumption.

Table 5: Encryption algorithm memory usage on micaz and telosb sensor motes [32]

| Encryption Algorithm | MicaZ | | TelosB | |
|---|---|---|---|---|
| | RAM (KB) | ROM (KB) | RAM (KB) | ROM (KB) |
| RC5 | 0.2 | 2.5 | 0.2 | 6 |
| AES | 2 | 10 | 1.8 | 9 |
| Skipjack | 0.6 | 10 | 0.04 | 7.5 |
| XXTEA | 0.049 | 3.1 | 0.04 | 3.8 |

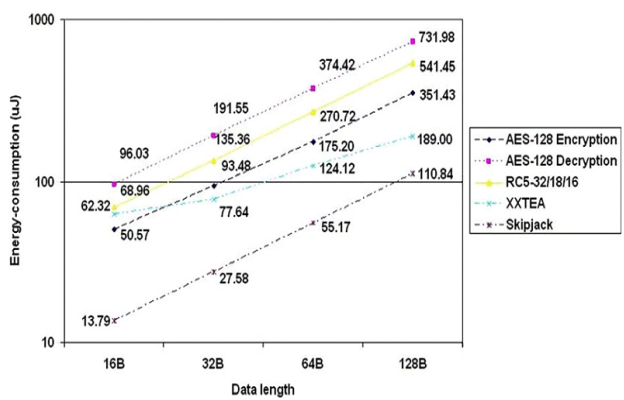

Fig. 2 Energy consumption of block cipher on Micaz sensor motes [32].

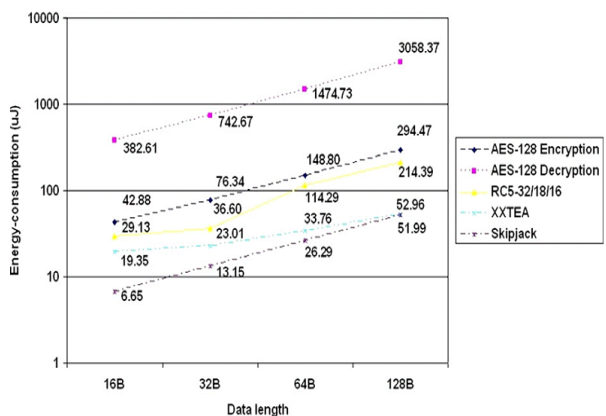

Fig. 3 Energy consumption of block cipher on TelosB sensor motes [32].

[32] Shows Skipjack and XXTEA have efficient energy consumption and smaller memory requirements while RC5 and AES provide better security. The energy consumption on RC5 and AES depends on the key size and number of rounds respectively. RC5 consumes more energy in its encryption phase than that of AES on MicaZ, but overall energy consumption and memory of RC5 is less. Similarly on TelosB, RC5 consume less energy for both encryption and decryption than AES and uses less memory.

RC5 provide good security against different attacks and is also a fast block cipher algorithm suitable for both hardware and software implementation. Since it is a parameterized algorithm having variable features (like block size, number of rounds and length secret key) it provides flexibility in both performance and security [31], [33].

### 3.3 Can we apply all given methods to create MAC like block cipher, hash function and stream cipher etc?

Yes we can, but we have to look to the resources and we need to choose such a technique which is feasible for sensor networks. In case of more resources the best option is to go for hash function because it provides better security. But according to the current situation, sensor network has limited resources, it is common practice to use block cipher for implementation of MAC in sensor networks as it requires less resources comparatively.

Thus, from the above discussion we conclude that block cipher is the best option to create MAC in WSNs.

## 4. Conclusion

In this paper we investigate that public key is not energy efficient and is expensive in terms of both computation and communication as compared to symmetric key. Sensor networks has limited resources, therefore most of the researcher used symmetric key to create MAC in WSNs. Thus, we conclude that symmetric key techniques are more feasible for WSNs as compared to public key.

After selecting symmetric key techniques, we compared different attacks on hashing techniques and conclude that it offers good security mechanisms as compared to block cipher. However, it requires an order of magnitude higher overhead and also uses more memory. While stream cipher, Skipjack and XXTEA are less secure than block cipher and for encryption packets overhead also takes place in stream cipher. By selecting an efficient technique, we pick block cipher as best technique to create Message authentication code (MAC) in sensor network although hash function offers good security.

We conclude that RC5 is feasible and consumes less energy/resources as compared to other algorithms (AES, MD5, SHA1, IDEA) except Skipjack and XXTEA. However, RC5 is more secure than Skipjack and XXTEA. Thus we propose that RC5 is a best algorithm to create MAC in sensor networks.

**First Author** Sadaqat Ur Rehman has completed his BS degree program in Computer System Engineering from N.W.F.P University of Engineering and Technology Peshawar, his areas of interest are Wireless sensor Networks, Artificial Intelligence and Microcontroller. Currently he is Lab Engineer in

Electrical Engineering Department at Sarhad University of Sciences and Information Technology Peshawar

**Second Author** Muhammad Bilal is a student of BS degree program in Computer System Engineering at N.W.F.P University of Engineering and Technology Peshawar, his areas of interest are Wireless sensor Networks, Digital Image Processing and Artificial Intelligence.

**Third Author** Basharat Ahmad is a student of BS degree program in Computer System Engineering at N.W.F.P University of Engineering and Technology Peshawar, his areas of interest are Wireless sensor Networks, Microcontroller and Data Base Management System.

**Fourth Author** Khawaja Muhammad Yahya has completed his MS degree in Computer Engineering from University of Missouri-Rolla, USA in 1987. He Completed PhD in Information Management System / Decision Support System from University of Missouri-Rolla, USA in 1995. Currently he is a Chairman of Department of computer System Engineering at N.W.F.P University of Engineering and Technology Peshawar.

**Fifth Author** Anees Ullah has completed his BS degree in Electrical Engineering from N.W.F.P University of Engineering and Technology Peshawar; currently he is pursuing MS degree from the same university.

**Sixth Author** Obaid Ur Rehman has completed his BS degree in Electrical Engineering from N.W.F.P University of Engineering and Technology Peshawar, MS degree from university of Liverpool UK. Currently he is

Assistant professor in Electrical Engineering Department at Sarhad University of Sciences and Information Technology Peshawar.